\documentstyle[aaspp]{article}
\def\etal{{\it et al. }}
\def\ltsim{\mathrel{\hbox{\rlap{\hbox{\lower4pt\hbox{$\sim$}}}\hbox{$<$}}}}
\def\gtsim{\mathrel{\hbox{\rlap{\hbox{\lower4pt\hbox{$\sim$}}}\hbox{$>$}}}}

\def\apj    {{\it Astrophys.~J.~}}
\def\apjl   {{\it Astrophys.~J.~(Letters)~}}
\def\apjs   {{\it Astrophys.~J.~Suppl.~}}

\def\aj     {{\it Astr.~J.~}}
\def\aa     {{\it Astr.~Ap.~}}
\def\mnras  {{\it Mon. Not. R. astr. Soc.~}}

\def\pasp   {{\it Pub.~Astr. Soc. Pac.~}}
\def\araa   {{\it Ann. Rev. Astr. Ap.~}}
\def\baas   {{\it Bulletin of the American Astronomical Society~}}

\begin{document}

\title{HST Imaging of the Globular Clusters in the Fornax Cluster: NGC 1379}

\author{Rebecca A. W. Elson}
\affil{Institute of Astronomy, Madingley Road, Cambridge CB3 0HA,
UK}
\affil{Electronic mail: elson@ast.cam.ac.uk}

\author{Carl J. Grillmair}
\affil{Jet Propulsion Laboratory, 4800 Oak Grove Drive, Pasadena, CA 91109 USA}
\affil{Electronic mail: carl@grandpa.jpl.nasa.gov}

\author{Duncan A. Forbes}
\affil{School of Physics and Astronomy, University of Birmingham,
Edgbaston, Birmingham B15 2TT, UK}
\affil{Electronic mail: forbes@star.sr.bham.ac.uk}

\author{Mike Rabban}
\affil{Lick Observatory, University of California, Santa Cruz, CA 95064 USA}
\affil{Electronic mail: mrabban@ucolick.org}

\author{Gerard. M. Williger}
\affil{Goddard Space Flight Center, Greenbelt, MD 20771 USA}
\affil{Electronic mail: williger@tejut.gsfc.nasa.gov}

\author{Jean P. Brodie}
\affil{Lick Observatory, University of California, Santa Cruz, CA 95064 USA}
\affil{Electronic mail: brodie@ucolick.org}

\begin{abstract}

We present $B$ and $I$ photometry for $\sim 300$ globular cluster
candidates in NGC 1379, an E0 galaxy in the Fornax Cluster.  Our data
are from both Hubble Space Telescope (HST) and ground-based observations. 
The HST photometry ($B$ only) is essentially 
complete and free of foreground/background
contamination to $\sim 2$ mag fainter than the peak of the globular
cluster luminosity function.
Fitting a Gaussian to the luminosity function we find $\langle B \rangle
=24.95\pm0.30$ 
and $\sigma_B = 1.55\pm0.21$.
We estimate the total number of globular clusters to be  
$436\pm 30$.  To a radius of 70 arcsec we derive a moderate specific
frequency, $S_N=3.5 \pm 0.4$.  
At radii $r\sim 3-6$ kpc
the surface density 
profile of the globular cluster system is indistinguishable 
from that of the underlying galaxy light.
At $r \ltsim 2.5$ kpc the profile of the globular cluster system flattens,
and at $r \ltsim 1$ kpc, the number density  appears to decrease.
The $(B-I)$ colour distribution of the globular clusters (from ground-based
data) 
is similar to that for Milky Way globulars, once corrected for background
contamination.  It shows no evidence for bimodality or for the presence
of a population with [Fe/H]$\gtsim -0.5$. 
Unlike in the
case of larger, centrally located cluster ellipticals, neither mergers
nor a multiphase collapse are required to explain the formation of
the NGC 1379 globular cluster system.

We stress the importance of corrections for background
contamination in ground-based samples of this kind:  
the area covered by a globular cluster system (with radius $\sim 30$ kpc)
at the distance of the Virgo or Fornax cluster
contains $\gtsim 200$ background galaxies 
unresolved from the ground, with magnitudes 
comparable to brighter 
globular clusters at that distance.  
The colour distribution of these galaxies is strongly peaked
slightly bluer than the peak of a typical 
globular cluster distribution.
Such contamination 
can thus create the impression of skewed
colour distributions, or even of bimodality, where none exists.


\end{abstract}


\section{Introduction}

An accumulating body of observations suggests that
the distribution
of colours of globular clusters,
and by inference of their metallicities,  varies significantly from 
one galaxy to the next.  
Of particular interest is the colour bimodality 
which has now been
observed in  the globular cluster systems of
several large elliptical galaxies, and 
suggests the presence of 
distinct metal rich and metal poor populations.  
The best example is the Virgo cD galaxy M87 (NGC 4486) 
(cf. Elson \& Santiago 1996).
It has one population of globular clusters with colours
similar to those of the Milky Way globulars, and one which is significantly
redder, with inferred metallicities $\gtsim$ solar. 
Other galaxies whose globular cluster systems 
show clear bimodality include M49 (NGC 4472), an E2 galaxy in the Virgo Cluster with
the same luminosity as M87 (Geisler, Lee \& Kim
1996), and NGC 5846, a slightly less luminous E0 galaxy 
at the centre of a small compact group (Forbes, Brodie \& Huchra 1997a).

Two ideas have been invoked to explain the colour bimodalities.
One is that elliptical galaxies are formed during
mergers in which populations of metal rich (red)
globulars are created and 
added to a `native' 
population of metal poor (blue) clusters (cf. Ashman \& Zepf 1992). 
The other is that
globular cluster populations with different mean metallicities
form during a multiphase collapse of a single
system (Forbes, Brodie \& Grillmair 1997b): metal poor globular clusters are
formed early in the collapse, 
while metal rich ones form later, roughly contemporaneously with the stars.

To understand fully  the implications of the observed colour distributions 
for the origin 
of globular cluster systems, a much larger body of accurate data for systems 
surrounding galaxies of a variety of types and in a variety of environments
is required.  A question of particular importance, for example, is whether
all elliptical galaxies have bimodal globular cluster systems, 
or whether such systems are restricted to large galaxies in rich 
environments.  
The data best suited to address this question are those
acquired with the Hubble Space Telescope (HST).  The resolution of HST
allows even  
the crowded central regions of
galaxies 
at the distance of Fornax and Virgo to be probed and allows most background
galaxies to be eliminated. 
At these distances samples of globular clusters
are complete and uncontaminated to well past the peak of 
the luminosity function.  
This paper and the others in this series 
are contributions to this growing database.  
Forbes \etal (1997a) and Grillmair \etal (1997) discuss the globular cluster
systems of the Fornax cD galaxy NGC 1399, its neighbour the E1
galaxy NGC 1404, and the peculiar galaxy NGC 1316 which may have undergone
a recent merger.  Here we
present observations of the 
globular cluster system of NGC 1379, a normal E0 galaxy in the Fornax
cluster.  

Hanes \& Harris (1986) used photographic data to study the 
NGC 1379 globular cluster system to $B=23.6$ (about a magnitude brighter
than the peak of the luminosity function).  They measured the 
profile of the outer part of the system ($5 < r < 35$ kpc), and estimated
the total population to number $\sim 800$.  
More recently  
the globular cluster systems of five
galaxies in the Fornax cluster, including NGC 1379, have been studied by Kohle \etal 
(1996) and Kissler-Patig \etal (1997a) using $V$ and $I-$band photometry
obtained with the 100-inch telescope at Las Campanas. 
Their data are 50\% complete at 
$B \sim$ 24, and cover a radial range $\sim 3 - 10$ kpc. 

Our observations, obtained both from the ground at the Cerro Tololo
Interamerican Observatory (CTIO),
and from space using HST,  
are described in 
Section 2.  Section 3 presents the results, including a caveat
concerning the need
for accurate background corrections 
for ground-based data.  Our findings are 
summarized in Section 4.  

\section{Observations and Data Reduction}

In this section we describe the HST and CTIO 
images upon which our results are based,
and the process for detecting, selecting, and determining magnitudes for
the globular cluster candidates in each case.  
We also discuss completeness, and contamination from foreground
stars and background galaxies.  
At an adopted distance of 18.4 Mpc ($m-M=31.32$; Madore \etal 1996), 
1 arcsec corresponds to 89 pc.  Reddening in the direction of the
Fornax cluster is assumed to be negligible (Bender, Burstein, \&
Faber 1992). 

\subsection{HST Imaging}

Five images of NGC 1379 were obtained with HST on 1996 March 11,
using the F450W ($\sim B$) 
filter.  (Due to technical difficulties, the complementary $I$-band
images were not acquired, and are anticipated in 1997.)
Three images were taken at one pointing, and two more were offset
by 0.5 arcsec.
The total exposure time was 5000 seconds.  
The centre of NGC 1379 was positioned at the centre of the Planetary
Camera (PC) chip to
afford the greatest resolution in the most crowded regions.
Details of the reductions
are given by Grillmair \etal (1997).  Briefly, 
the images were reduced using the standard pipeline procedure.
The VISTA routine SNUC was used to fit and subtract the underlying galaxy.
We ran
DAOPHOT II/ALLSTAR (Stetson 1987)  
separately on the sum of the first three images and the last two images,
requiring that detections appear in both lists to qualify as real
objects.
We adopted a detection threshold of 3$\sigma$ and measured magnitudes by
fitting a point-spread function (PSF).
Extended objects were eliminated by visual inspection.  
Count rates were converted to $B$
magnitudes using the gain ratios and zeropoints given by Holtzman \etal
(1995; 1997, private communication). Photometry is available on
request from CJG.

Figure 1 shows a mosaic
of the four WFPC2 chips, with the galaxy subtracted.
The total area of the field, excluding two 60 pixel wide unexposed borders
on each chip, is 4.8 arcmin$^2$.  The scale of the Wide Field Camera
(WFC) is 0.0996 
arcsec pixel$^{-1}$,
and of the PC, 0.0455 arcsec pixel$^{-1}$.  At the distance of NGC 1379, 
one WFC pixel corresponds to $\sim 9$ pc and one PC pixel to $\sim 4$ pc. 
A globular cluster with size typical of those in the Milky Way (core radius 
$\sim 2$ pc, half-mass radius $\sim 10$ pc, and tidal radius $\sim 50$ pc)
will thus
appear essentially unresolved in our images.

A total of $\sim 300$ objects were detected and measured in our field.
To determine the completeness of this sample, 3000 artificial PSFs
(100 at each of 30 magnitude levels)
were added to the images, and the images were then processed in a manner
identical to that
for the original data.
The completeness of the sample as a function of magnitude is shown
in Figure 2.  The sample is $\sim 100$\% complete to $B=26$ for the WFC 
chips (80\% of the sample), and to $B=25.5$ for the PC chip.
At $B > 26$ the completeness begins to drop rapidly.
Photometric errors are  
$\delta B \approx 0.10$ mag at $B \ltsim 25$, rising to 0.15 mag at 
$B=26$.

Next we consider the extent to which our sample may be contaminated
by foreground stars and  background galaxies.  Few foreground stars are 
expected in an area of only 4.8
arcmin$^2$ at this Galactic latitude, and
most background galaxies are resolved and thus easily distinguished from 
globular clusters.  The main source of contamination is 
compact,
spherical background galaxies.  To determine the expected 
level of contamination in our sample, we observed a background field located
$\sim 1.4$ degrees south of the center of the Fornax cluster.
The exposure time was 5200 seconds, so the limit of detection is comparable
to that for the NGC 1379 sample.  The image was processed and the 
sample selected in the same
way as for the NGC 1379 field (see Grillmair \etal 1997). 

Figure 3 shows a colour-magnitude diagram (CMD) for the 84 unresolved objects
detected in the background field.  The sample becomes incomplete at 
$B > 26.5$, but as we shall see, this is $\sim 1.5$ magnitudes fainter
than the peak of the luminosity function, and so will not affect our
results.  At $B \ltsim 26.5$ the objects have a wide range of colours,
with the majority concentrated around $(B-I)\sim 1$.
The $B$ luminosity function for the background sample
is plotted in Fig. 4, which also shows the luminosity function for the
$\sim 300$ candidate globular clusters.  The background luminosity\
function is tabulated in Table 1.  Since these
background number counts are  applicable to HST studies of
any unresolved population at high latitude, we also
include in Table 1 the $I$-band luminosity function for the background field.
This is plotted as the solid histogram in Fig. 5. 
As a check on the consistency of the sample, and to assess the amplitude of 
spatial fluctuations in the  background on this scale, 
we compared the luminosity
function in Fig. 5  with the
Medium Deep Survey (MDS) star count data from 17 high latitude fields
obtained with HST in the $V$ and $I$ bands (Santiago, Gilmore \& Elson 1996).
The dotted histogram in Fig. 5 
shows the $I$-band luminosity function for the MDS data, 
normalized to an area of 4.8 arcmin$^2$.  The two distributions are in excellent 
agreement  to $I\approx 23.5$ which is 
the limiting magitude of the MDS star count
data.   Fainter than this stars can no longer be reliably distinguished
from compact galaxies.  The MDS stellar luminosity
function in $V$, normalized to the same area, is also included in Table 1.
Finally, Fig. 5 also shows a luminosity function for stars and galaxies from
the Canada-France Redshift Survey (CFRS) (Lilly \etal 1995; S. Lilly 1997,
private communication).  The magnitudes were measured in an aperture of
diameter 3 arcsec.  The $I$-band data cover an effective area of $\sim 425$
arcmin$^2$, and have again been normalized to an area of 4.8 arcmin$^2$.
These data illustrate the much larger degree
of background contamination to be expected in the extreme case where
no galaxies (with $I \gtsim 21$)
can be distinguished morphologically from unresolved objects.
Background contamination in typical ground-based samples of globular
clusters in distant galaxies will fall somewhere between
the CFRS and the HST histograms.

\subsection{Ground--based Imaging}

Broadband $B$ and $I$ images of NGC 1379 were taken with the 
CTIO 1.5m telescope, 
using a Tek 2048 x 2048 array with a pixel
scale of 0.44 arcsec pixel$^{-1}$.  Total exposure times were 9000 and 3900
seconds for the $B$ and $I$ images respectively.
Although the seeing
was only $\sim$ 1.5 arcsec, conditions remained photometric throughout the
night of 1995 December 24. Reduction was carried out in the standard
way (i.e. bias and
dark subtraction, flat--fielding and sky subtraction). 
After combining, the images were
calibrated using aperture photometry from the catalogs of 
of Longo \& de Vaucouleurs (1983) and de Vaucouleurs \& Longo (1988). 
This procedure gave an rms precision of
better than 0.05 mag. 

For the selection of globular cluster candidates in the CTIO images, we 
employed an iterative procedure using DAOPHOT II. 
We measured the background
noise in both images and set the threshold for single pixel detection 
at 5$\sigma$,
i.e. five times the noise due to the background. The other important detection
parameters, SHARPness and ROUNDness (designed to weed out extended
objects and cosmic rays), were initially given a
large range. For each detected object we measured a 3 pixel radius
aperture magnitude and applied an aperture correction based on a
curve-of-growth analysis for a dozen isolated globular clusters. 
The rms error in 
the aperture correction is $\sim$ 0.05 mag. 

We then
compared our $B$-band candidate list with the positions and $B$
magnitudes of globular clusters  
detected with HST's WFC. 
Our candidate list was matched to the HST list with the condition
that the HST globular cluster
lie within 3 CTIO pixels of our object. With this
condition, 28 objects were matched. 
The average magnitude difference
between the CTIO and HST $B$ magnitude is 0.03 mag. 
Such excellent agreement is gratifying
from a photometric standpoint, and reassures us that we have matched
the data sets correctly.
The matched
clusters have SHARPness parameters 0.3 to 0.7 and ROUNDness $-0.4$  
to 0.4. 
Assuming that these globular clusters 
are representative of all globular clusters in the CTIO image, we re--ran
DAOPHOT II with the new restricted range in SHARPness and ROUNDness
parameters.  This resulted in the exclusion of 
about half the objects in the original sample, 
which are presumably background galaxies.
The same parameters were used for both the $B$ and $I$ images. 
Figure 6 shows a CMD for 365 objects selected
in this way.

A comparison of the CTIO object list and the HST
list, within the area in common, will also indicate how complete our
detection is as a function of magnitude. The 
completeness function estimated this way is shown in Fig. 7. Our sample is 
$\sim 100$\% complete at  $B \sim$ 21 and $\sim 50$\% complete at 
$B \sim$ 23.5. 
The photometric errors are $< 0.1$ mag
for all magnitudes brighter than our 50\% completeness limit.

At this point it is necessary to investigate the level of contamination
of our ground-based 
sample by foreground stars and background galaxies.  Since the
field is much larger than that observed with HST 
(211 arcmin$^2$ compared to 
4.8 arcmin$^2$), and since the resolution is not sufficient to distinguish
many background galaxies from unresolved sources, contamination from
both stars and galaxies will be significant.  
Since no background comparison field
was obtained, we rely instead on the CFRS
data in the $B$ and $I$-bands, discussed above, to estimate a
correction for contamination. 

Figure 8 shows histograms of $(B-I)$ for the full NGC 1379 sample in 
Fig. 6, 
and for the CFRS sample of objects with $19 < B < 23$.  Since
selection effects in the two samples are different, we do not normalize the
CFRS sample directly using the known area of the survey.  Rather, we
normalize it to match the tail of objects with $(B-I) > 2.5$ in Fig. 8, 
whose colours are too red to be globular
clusters.  From this we can infer the relative number of contaminants with 
$(B-I) < 2.5$.
The CFRS sample is
strongly peaked at $(B-I) \sim 1.0$, which is just  
$\sim 0.4$ mag bluer than the peak of the 
NGC 1379 sample.  The NGC 1379 sample is lacking many of the bluest
objects in the CFRS sample;  this is probably because fainter 
galaxies are on average bluer,
and the CFRS sample is much more complete than the
NGC 1379 sample at the faintest magnitudes.  Also, our process of 
excluding galaxies from the CTIO sample on the basis of ROUNDness
would preferentially eliminate edge-on spirals,
which are systematically bluer than ellipticals. 
The most important point
illustrated by Fig. 8 is that in the ground-based data, much more
so than in
the HST data, failure to properly correct for foreground/background 
contamination may lead to a significant error in the deduced colour of
the peak of the globular cluster
colour distribution, and possibly to the erroneous impression of a 
bimodal colour distribution.  With ground-based data obtained 
with better seeing it would of course be possible to  exclude a 
greater proportion of background galaxies, so that any skewing of the
colour distribution would be less severe. 

\section{Properties of the NGC 1379 globular cluster system}

The principal properties of a globular cluster system which any theory
of its origin and evolution must account for are 
the luminosity 
function (in particular, the absolute magnitude of its peak, and the width
of the distribution), the colour distribution (in particular, the colour
of the peak and the presence or absence of any bimodality), the
radial distribution, and the total number of clusters, $N_{tot}$, 
and therefore
the specific frequency, defined as $S_N=N_{tot} 10^{0.4(M_V+15)}$,  
where $M_V$ is the absolute magnitude of the galaxy.
With $B_{tot}=12.07$ (Tully 1988) and  
$(B-V)=0.89$ (Faber \etal 1989) we infer an apparent magnitude for NGC 1379
of $V_{tot}=11.18$.  With distance modulus
31.32, this implies $M_V=-20.16$.  This is the integrated magnitude 
within a radius of $\sim 70$ arcsec.

\subsection{Luminosity function}

The luminosity function for the NGC 1379 globular cluster candidates
derived from 
the HST data is shown in Fig. 4.  This sample includes $\sim 300$
unresolved objects and is  
essentially complete and uncontaminated well past the
peak. The background-corrected luminosity function, that is,
the difference between the solid and dashed histograms in Fig. 4,
is shown in Fig. 9.
Fitting a Gaussian function to the 
luminosity function in Fig. 4, using a maximum-likelihood technique
which is independent of binning, and 
takes into account completeness, background contamination, and 
photometric errors (Secker \& Harris 1993),
we derive a peak
magnitude of $\langle B \rangle =24.95 \pm 0.30$, and a width
$\sigma_B=1.55 \pm 0.21$.   This Gaussian is shown as the solid
curve in Fig. 9.

We know of no other direct measurements of $\langle B \rangle$ for this
system to compare with ours.  
However, our values are the same, to within the errors, as the values
measured for the globular cluster systems of two other Fornax 
galaxies, NGC 1399 and NGC 1404
(Grillmair \etal 1997).
Also,  
for four elliptical galaxies in the Virgo cluster Harris
\etal (1991) find an average value 
$\langle B \rangle=24.77 \pm 0.2$.  Combining this
with the value  
$(m-M)_{Fornax} - (m-M)_{Virgo} = 0.08 \pm 0.09$ (Kohle \etal 1996) 
implies
$\langle B \rangle=24.85\pm 0.22 $ for NGC 1379.  This is 
in good agreement with our value.  
  
We may either use the measured peak magnitude to infer a  
distance modulus for NGC 1379, assuming a `universal' value for the absolute
magnitude of the peak, or we may adopt a distance modulus and infer
an absolute magnitude.
While the value of $\langle M_V \rangle$ has been measured for many globular
cluster systems, there are fewer $B-$band studies. 
Sandage \& Tamman (1995) quote $\langle M_B \rangle = -6.93 \pm 0.08$ for the 
Milky Way and M31 globular
clusters, while  Ashman, Conti \& Zepf (1995) give $\langle M_B \rangle = -6.50$
for the Milky Way clusters.  
Adopting the Cepheid distance for NGC 1379 and our measured peak magnitude,
implies
$\langle M_B \rangle = -6.37 \pm0.36$, which is in good agreement with
the Ashman \etal value and somewhat fainter than the
Sandage \& Tamman value.

As we shall see below, 
the $(B-I)$ distribution of NGC 1379 globular clusters appears
very similar to that for the Milky Way globular clusters. It should 
therefore be safe to
assume that the $(B-V)$ distribution is also similar.  Adopting
$\langle B-V \rangle = 0.7$ for the Milky Way globulars (Harris 1996),
we may convert our value for
the peak magnitude of the luminosity function, $\langle B \rangle = 24.95
\pm 0.30$, to $\langle V \rangle = 24.25\pm0.30$.
This value is somewhat fainter than
the value found by Kohle \etal (1996)
of $\langle V \rangle = 23.68 \pm 0.28$, but
their data reach just to the peak of the luminosity function, and the
errors in their faintest bin are large.
The sense of the difference is consistent with our result above that
fitting only the brighter part of the luminosity function results in
a peak magnitude which is too bright.
With the Cepheid distance modulus of 31.32, our $\langle V \rangle$ value
implies $\langle M_V \rangle = -7.07 \pm 0.36$, which
is typical for elliptical galaxies of this absolute
magnitude (Harris 1991).

Our value of $\sigma_B=1.55\pm0.21$ for the NGC 1379 globular cluster system is 
larger than the values $\sigma_B = 1.07$ and 0.89 quoted by Sandage \&
Tamman for the Milky Way and M31 respectively, 
but is consistent with the 
values $\sigma_B=1.37\pm 0.07$ and
$\sigma_B=1.39\pm 0.12$ found by Grillmair \etal (1997) for the globular
cluster systems of NGC 1399 and NGC 1404 respectively.
It is also consistent with the 
value $\sigma_B=1.46\pm 0.07$ 
for four elliptical galaxies in the Virgo cluster (Harris \etal 1991).

\subsection{Radial profile}

Harris \& Hanes (1987) compared the radial profile of the NGC 1379 globular
cluster system with the surface brightness profile of the galaxy itself from 
Schombert (1986), over the radial range $5-35$ kpc.  They detected no
difference between the two, although their uncertainties were large.
Kissler-Patig \etal (1997a) found the same result from their ground-based
data over the range $3-10$ kpc.  
Radial surface density 
profiles for our HST sample of NGC 1379 globulars
are shown in Figs. 10a and b.
Corrections have been made to compensate for the fraction of each
annulus which falls outside the field of view of the WFPC2. In Figure 10a
profiles are plotted for both the complete, uncontaminated 
sample with $B < 25.5$, and for the objects with $B > 26.5$ (with no
correction for completeness), which
are expected to be primarily background galaxies.  
Indeed, the fainter
sample shows almost no radial gradient.  The dashed line is the background
level ($\sim 9.0$ objects per arcmin$^2$)
measured from the background field for $B > 26.5$.  The
agreement is excellent.  The background level for the brighter sample,
again measured from the background field, is 2.7 objects per arcmin$^2$,
and is shown as the solid line.  The difference between the radial profile
and the background level is probably due to small numbers, but may 
indicate a small amount of residual contamination:  the five
outermost points represent an average of $< 4$ objects each, and the 
total number of objects in the background field with $B < 25.5$ is just 13.

The radial profile of the 
brighter sample decreases smoothly from $\sim 10-80$ arcsec  
($\sim 1-7$ kpc), at which point the globular cluster system is lost in the 
background. 
Figure 10b shows the radial profile for the sample with $B < 25.5$,
with a background of 2.7 objects per arcmin$^2$ subtracted.
Superposed is a curve representing the surface brightness
profile of the underlying galaxy from Kissler-Patig \etal (1997a), scaled
arbitrarily to match the profile of the globular cluster system.
The two profiles agree well at $\sim 35-70$ arcsec (3--6 kpc).
There is no evidence that the surface brightness
profile of the globular cluster system is shallower than that of the
underlying galaxy light out at least
to the limit of our data at $r\sim 7$ kpc.
The logarithmic slope of the profile in Fig. 10b is $-2.4$ at 
$ r \gtsim  35$ arcsec.
 
Inwards of $r \sim 30$ arcsec ($\sim 2.5$ kpc) the profile of the globular cluster
system flattens out.
This  core structure 
seems to be a common feature of the globular cluster
systems of elliptical galaxies, and the radius at which the flattening 
occurs correlates with the galaxy luminosity (Forbes \etal 1996).
The mean surface density within $\sim 10$ arcsec of the centre of NGC 1379 is 
$\sim 200\pm60$ clusters per arcmin$^2$.
The core radius, where the surface density has fallen to half its
central value, is $r_c \sim 23\pm 6$ arcsec ($2.0\pm0.5$ kpc).  This is 
consistent with values 
for other  galaxies with absolute
magnitude comparable to that of 
NGC 1379.  
Such a core structure is not present in the 
underlying galaxy light, which, while it changes slope slightly 
at $r\sim 50$ 
arcsec, rises
with constant slope inwards to at least 10 arcsec (Schombert 1986).

Inwards of $\sim 10$ arcsec
($\sim 1$ kpc), the surface density of globular clusters appears to decrease.
This radius corresponds to 220 pixels on the PC,
and as can be seen from Fig. 1, crowding even in these inner
regions is not severe.  Extrapolating a smoothly rising profile to
the center of the galaxy would require 6 clusters instead of 2 in
the innermost bin, and 14 instead of 12 in the second bin.  If the
radial distribution really were steadily rising, this would imply that
our data are only $\sim 70$\% complete in these combined bins.  
Closer analysis of our
completeness tests reveals that we are in fact 97\% complete for $B <
25.5$ in the region $r < 220$ pixels (the slight reduction over the
PC-averaged completeness value being a consequence of the increased
noise due to the integrated stellar light of the central regions of
the galaxy). We conclude that the central dip in the cluster surface
density distribution is not an artifact of our analysis, and may
indeed indicate a quite substantial drop in the volume density of
clusters near the nucleus of the galaxy. Radii less than 1 kpc from
the nucleus are where we expect tidal stresses to begin to take a toll
on the numbers of globular clusters (Lauer \& Kormendy 1986,
Grillmair, Pritchet, \& van den Bergh 1986), either during their
formation or subsequently through tidal stripping (Grillmair \etal
1995), particularly if the clusters are on box orbits.

We now turn to the ground-based data which cover a much 
greater area than the HST data.
The radial distribution for the full sample from Fig. 6 
is shown in Fig. 11.  The surface density for the 
CTIO sample has been increased by 0.74 in  
$\log N$ 
to match the HST sample, which is shown as the solid curve.  
At $r\gtsim 100$ arcsec, the profile of the CTIO sample 
is essentially flat, suggesting that the sample is composed overwhelmingly
of foreground/background objects at these radii. 
Even a colour selection
is unlikely to help disentangle the background contamination
since, as shown in Fig. 8, 
background galaxies have a 
similar  colour distribution to the full NGC 1379 sample.  We therefore
conclude that, despite the larger spatial coverage of the ground-based
sample, in the absence of a suitable background calibration field 
it does not contribute much to our knowledge of the
radial structure of the NGC 1379 globular cluster system not covered
by our HST data.  It is also
unable to probe the innermost regions of the cluster system, due to 
crowding. 

\subsection{Colour distribution}

In the absence of an $I$-band image from HST, we 
attempt to extract  
a sample of globular clusters 
from the ground-based data  
which is as uncontaminated as possible,  
to investigate the $(B-I)$ colour distribution.
One approach is to select
only objects with 
$r < 100$ arcsec, since these 
show a radial gradient in surface density (Fig. 11).  
This gives a sub-sample of 35 objects.  The background level inferred from 
the radial distribution of the CTIO sample at $r > 100$ arcsec is 1.6 
arcmin$^{-2}$, implying that 14 of the 35 globular cluster candidates
at $r < 100$ arcsec, 
or 40\% of this sample, are background galaxies or foreground stars.

The colour distribution for the 35 objects at $r < 100$ arcsec, along with 
those for the full CTIO sample and for the CFRS sample, is shown as 
the dashed histogram in Fig. 8.
The $r < 100$ arcsec sample clearly peaks at a redder colour than the
full sample, underlining the fact that the colour
distribution of the uncorrected
sample is misleading.  
Figure 12 shows the same  histogram for the $r < 100$ arcsec 
sample, along with the histogram of residuals obtained by subtracting
the normalized CFRS histogram from the full CTIO histogram in Fig. 8.
A histogram for 95 globular clusters in the Milky Way (Harris 1996)
is also shown.  The colour distributions of the NGC 1379 sample with
$r < 100$ arcsec, and with the CFRS sample subtracted, have a very similar
peak colour, $(B-I)\approx 1.6$, which is  indistinguishable from that for the
globular cluster system of the Milky Way. 

Using the relation between $(B-I)$ colour and metallicity from 
(Couture \etal 1990) we infer from  the peak colour 
a metallicity of [Fe/H]$\sim -1.5$. 
Forbes \etal (1997b) plot a relation between mean metallicity of 
globular clusters and
parent galaxy magnitude for 11 galaxies with $-21 < M_V < 23$,
 with bimodal globular
cluster colour distributions.  They find that,
while in the metal rich populations ([Fe/H]$> -0.5$) there is a strong
correlation between mean globular cluster
metallicity and galaxy magnitude, for the metal poor populations 
the scatter is much greater and the correlation much weaker.
There is, however, a trend for less luminous galaxies to have 
globular clusters with a lower mean metallicity, and our results for
NGC 1379 are consistent with this trend.

There is a tail of bluer objects
in the NGC 1379 samples which is probably comprised of residual
background galaxies.  If these objects were globular clusters,
their colours
would imply metallicities of
[Fe/H]$\ltsim -2.5$.  The objects with  $(B-I)\gtsim 2.5$
are again probably residual
stars or galaxies.  Even the most metal rich globular clusters in
the Milky Way have $(B-I) \ltsim 2.0$, and a metal rich
population of clusters, such as appears to exist in M87, would
have $(B-I)\sim 2$.
We see no evidence for the presence of a
population of clusters with [Fe/H]$\gtsim -0.5$ in NGC 1379, nor for a bimodal
distribution of metallicities, although confirmation of this result
must await the acquisition of the complementary $I$-band image of our
field with HST.

It is interesting to compare the mean colour of the NGC 1379 globular clusters
with the colour of the underlying galaxy.  Pickles \& Visvanathan (1985)
present multicolour photometry for a sample of Fornax galaxies including
NGC 1379.  Their results imply $(B-I)\sim 2.0$ in an aperture of
diameter 45 arcsec.  This is significantly redder than the mean colour
of the clusters, and indicates that, as has been found in other galaxies 
(cf. Harris 1991),
the globular clusters of NGC 1379 are significantly more
metal poor than the halo stars.

\subsection{Total number and specific frequency}

We can estimate the total size of the NGC 1379 globular cluster 
population from our HST data.
Considering only globular clusters in the bright half of the luminosity
function (with $B < 24.95$), and making corrections for
the area of
each annulus that falls outside the WFPC2 field of view (typically $\sim 50-60 
$\%), 
we estimate that to $r=100$
arcsec, there are 218 globular clusters.  Multiplying by two to include 
the faint half of the luminosity function, we infer
a total number $N_{tot}=436\pm30$.  
Beyond 100 arcsec the globular cluster system is lost in the background. 
Since
in an annulus at $100-200$ arcsec we would expect $\sim 50$ background objects,
the number of globular clusters at these radii that have gone
undetected must be a small
fraction of 50.

Our estimate of $N_{tot}$ is consistent 
with the value
380$\pm 100$ quoted by Harris (1996), 
and is 40\% larger than the value $314\pm63$ estimated  
by Kissler-Patig \etal (1997a).
Our HST observations of the inner regions of the globular cluster system have
reduced the uncertainty in $N_{tot}$ considerably over estimates from
ground-based studies.
To $r=70$ arcsec we estimate $N_{r<70}=396 \pm44$.
Adopting $V_{tot}=11.18 $   
for NGC 1379 from Tully (1988), which is the integrated magnitude within
$r = 70$ arcsec, and the distance modulus 31.32,
we derive a 
specifc frequency $S_N=3.5\pm0.4$.  
This value is typical for E/S0 galaxies of this
magnitude (Harris 1991; Forbes \etal 1997b).

\section{Summary}

The properties of the NGC 1379 globular cluster system
derived in our study are summarized in Table 2.  In its luminosity
function and specific frequency, it appears to be typical
of globular cluster systems in elliptical galaxies of this magnitude.
The colour distribution of the globular clusters in NGC 1379 is 
similar to that of the Milky Way globular clusters.  In particular,
there is no evidence for bimodality or for the presence of a population of
clusters with [Fe/H]$\gtsim -0.5$.
Structurally, the outer part of the cluster system ($r > 3$ kpc) 
has the same surface density profile as 
the underlying galaxy light,
as suggested by previous studies 
(Harris \& Hanes 1987; Kissler-Patig \etal 1997a).

The similarity of the profiles of the  galaxy and globular cluster system
is not consistent with the predictions of merger models (cf. Ashman
\& Zepf 1992), which suggest that 
globular clusters formed during mergers are dynamically heated,
and the resulting 
system has a profile which is shallower than that of the underlying galaxy.
Neither is a multiphase collapse model required to explain these observations,
as there is no evidence of a second phase of globular 
cluster formation from 
enriched gas.
Indeed, NGC 1379 appears to be an example of an elliptical galaxy
whose globular clusters 
formed {\it in situ} in a single phase collapse, unlike the larger
centrally located ellipticals, including M87, NGC 1399, and NGC 5846,
which may require either mergers or a multiphase collapse to explain
their globular cluster systems.  
This result supports the suggestion of 
Kissler-Patig \etal (1997b) that low luminosity elliptical
galaxies ($M_V \gtsim -21.5$) have globular cluster
systems that formed early in a single collapse, and have remained
essentially unperturbed. 

Finally, we stress the importance of applying background corrections
to data obtained from the ground.  
At the distance of the Fornax or Virgo cluster, 
a globular cluster system with radius 30 kpc will cover an area that 
will contain $\sim 200$ background galaxies with $B \ltsim 25$, 
unresolved even with HST. 
A field with radius 10 kpc will contain $\sim 20$ such
objects.
The colour distribution of the background galaxies
is strongly peaked at 
$(B-I) \sim 1.0$.
Superposed on a distribution of somewhat redder
globular clusters, 
this can create the
impression of a skewed distribution or even of a bimodal one, where none
exists. The background galaxies would mimic a metal poor
(blue) globular cluster population rather than a metal rich (red) one.
On the other hand, an HST WFPC2 field with size $\sim 4.8$ 
arcmin$^2$ will generally contain
negligible numbers of foreground stars and background galaxies 
to $B\sim 26$.  While corrections are desirable in either case, they are
important in the case of ground-based observations of sparser globular
cluster systems, the results of which
may otherwise be misleading.
 
\noindent
{\bf Acknowledgments}\\
This research was funded in part 
by the HST grant GO-05990.01-94A and by Faculty
Research funds from UCSC.\\

\bigskip

\noindent{\bf References}

Ashman, K. \& Zepf, S. 1992 \apj 384, 50

Bender, R.,  Burstein, D.,  \& Faber, S. M. 1992 \apj 399 462

Couture, J., Harris, W. \& Allwright, J. 1990 \apjs 73, 671

Elson, R. A. W. \& Santiago, B. X. 1996 \mnras 280, 971

Faber, S. M., Wegner, G., Burstein, D., Davies, R. L., Dressler, A.,
Lynden-Bell, D. \&  Terlevich, R. J. 1989 \apjs 69, 763

Forbes, D. A., Brodie, J. \&  Huchra, J. 1997a \aj 113, 887

Forbes, D., Franx, M., Illingworth, G. \& Carollo, C. 1996 \apj 467, 126

Forbes, D. A., Brodie, J. \& Grillmair, C. G.  1997b \aj in press 

Forbes, D. A., Grillmair, C. G., Williger, G., Elson, R. \& Brodie, J., 
1997a, \mnras  in press

Geisler, D., Lee, M. G. \& Kim, E. 1996 \aj 111, 1529

Grillmair, C. J., Forbes, D. A., Brodie, J. \& Elson, R. A. W. 1997, \aj submitted 

Grillmair, C. J., Freeman, C. J., Irwin, M., \& Quinn, P. J. 1995,
\aj, 109, 2553

Grillmair, C. J., Pritchet, C. J., \& van den Bergh, S. 1986, \aj, 91,
1328

Hanes, D. A. \& Harris, W. E. 1986 \apj 309, 564

Harris, W. E. 1991 \araa 29, 543

Harris, W. E. 1996 \aj 112, 1487

Harris, W. E., Allwright, W. Pritchet, C. \& van den Bergh, S. 1991 \apjs 76, 115

Harris, W. E. \& Hanes, D. A. 1987 \aj 93, 1368

Holtzman, J. \etal 1995 \pasp 107, 1065

Kissler-Patig, M., Kohle, S., Hilker, M., Richtler, T., Infante, L.,
\& Quintana, H.  1997b \aa 319, 83

Kissler-Patig, M., Kohle, S., Hilker, M., Richtler, T., Infante, L.,
\& Quintana, H. 1997a \aa 319, 470

Kohle, S., Kissler-Patig, M., Hilker, M., Richtler, T., Infante, L., 
\& Quintana, H. 1996 \aa 309, L39

Lauer, T. R., \& Kormendy, J. 1986, \apj, 301, L1

Lilly, S., Le Fevre, O., Crampton, D., Hammer, F. \& Tresse, L. 1995 \apj 455, 50

Longo, G., \& de Vaucouleurs, A. 1983, {\it A General Catalogue of
Photometric Magnitudes and Colors in the UBV system} (University of
Texas, Austin)

Madore, B. \etal 1996, \baas 189, 108.04

Pickles, A. \& Visvanathan, V. 1985 \apj 294, 134

Santiago, B. X., Gilmore, G. \& Elson, R. A. W.  1996 \mnras 281, 871

Schombert, J. M. 1986 \apjs 60, 603

Secker, J. \& Harris, W. 1993 \aj 105, 1358

Stetson, P. B. 1987 \pasp 99, 191

Tully, B. 1988 {\it Nearby Galaxies Catalog}, Cambridge University Press, Cambridge

de Vaucouleurs, A. \&  Longo, G. 1988 {\it A Catalogue of Visual and Infra-Red
Photometry of Galaxies from $0.5\mu$m to $10 \mu$m}, Univ. of Texas, Austin

\vfill\eject

\begin{table}[t]
\caption{Luminosity functions for unresolved objects in high Galactic latitude
HST fields
.\label{tab:lum}}
\vspace{0.4cm}
\begin{tabular}{|c|c|c|c|c|c|c|c|c|c|}
\hline
$B$ & $N_{bck}$&& $I$ & $N_{bck}$ & $N_{MDS}$ && $V$ & $N_{MDS}$ \\
\hline

  20.25 &    0 &&   18.25 &    0 & 0    && 19.25 & 0 \\
  20.75 &    0 &&   18.75 &    1 & 0.18 && 19.75  & 0.13\\
  21.25 &    0 &&   19.25 &    1 & 0.76 && 20.25 & 0.19 \\
  21.75 &    0 &&   19.75 &    0 & 1.41 && 20.75 & 0.36 \\
  22.25 &    0 &&   20.25 &    0 & 1.47 && 21.25 & 0.85 \\
  22.75 &    1 &&   20.75 &    0 & 1.88 && 21.75 & 0.74 \\
  23.25 &    2 &&   21.25 &    2 & 1.82 && 22.25 & 1.65\\
  23.75 &    1 &&   21.75 &    2 & 2.35 && 22.75 & 1.53\\
  24.25 &    0 &&   22.25 &    1 & 2.47 && 23.25 & 2.27 \\
  24.75 &    3 &&   22.75 &    5 & 2.24 && 23.75 & 2.02\\
  25.25 &    6 &&   23.25 &    2 & 3.00 && 24.25 & 2.02 \\
  25.75 &    8 &&   23.75 &    5 & 1.18 && 24.75 & 2.08\\
  26.25 &   19 &&   24.25 &    6 & 0.41 && 25.25 & 2.51\\
  26.75 &   13 &&   24.75 &   11 & 0 && 25.75 & 1.78 \\
  27.25 &   12 &&   25.25 &   11 & &  && \\
  27.75 &   13 &&   25.75 &   16 & &  && \\
  28.25 &    6 &&   26.25 &   17 & &  && \\
  28.75 &    0 &&   26.75 &    4 & &  && \\
  29.25 &    0 &&   27.25 &    0 & &  && \\
  29.75 &    0 &&   27.75 &    0 & &  && \\

\hline
\end{tabular}
\vspace{0.4cm}

Notes:  Magnitudes are in the Johnson-Cousins system.
Columns 2 and 4 are numbers of unresolved objects (stars and 
galaxies) per unit magnitude in
our HST background field with area 4.8 arcmin$^2$, 1.4 degrees south of
the centre of the Fornax cluster.
Columns 5 and 7 are star counts from 17 high latitude HST fields observed
as part of the Medium Deep Survey.  Numbers are per unit magnitude, 
and are normalized to the area of a single
WFPC2 field (4.8 arcmin$^2$).

\end{table}

\begin{table}[t]
\caption{Properties of the NGC 1379 globular cluster system.\label{tab:pro}}
\vspace{0.4cm}
\begin{tabular}{|c|c|c|}
\hline
$\langle B \rangle =24.95\pm0.30$ & $N_{tot}=436\pm 30$ & $\langle M_B \rangle= -6.37\pm0.36 $ \\
$\sigma_B=1.55\pm0.21$ & $r_{core}=2.0\pm0.5$ kpc & $\langle M_V\rangle = -7.07 \pm$0.36 \\
$\langle B-I\rangle \approx 1.6$ & $\alpha = -2.4$ & $S_N=3.5 \pm 0.4$ \\
 &  &   [Fe/H]$\approx -1.5$ \\
\hline
\end{tabular}

\vspace{0.4cm}
Notes: Properties in columns 1 and 2 are derived directly from our
data. Properties in
column 3 are inferred assuming a distance modulus of 31.32, a colour for
globular clusters of $\langle B-V \rangle = 0.7$, a total magnitude for
NGC 1379 of $M_V=-20.16$, and a relation between $(B-I)$ and [Fe/H] from
Couture \etal (1990).  No correction for Galactic extinction
has been applied to any of 
these quantities, as it is assumed to be negligible.
The value of $S_N$ is for $r \ltsim 70$ arcsec.
\end{table}

\vfill\eject
\noindent
{\bf Figure Captions}

\noindent
{\bf Figure 1.} F450W($\sim B$) image of the NGC 1379 globular cluster 
system.
The underlying galaxy, centred on the smaller PC chip,
 has been subtracted.  The mosaic is 147 arcsec on a
side.

\noindent
{\bf
Figure 2.}  Completeness as a function of
$B$ magnitude for the HST sample, for the  WFC chips (solid curves)
and the PC
chip (dashed curve).  Our HST sample is $\sim 100$\% complete to
$B\sim 26$ for the WFC chips which contain $\sim 80$\% of the clusters,
and to $B\sim 25.5$ for the PC chip.

\noindent
{\bf
Figure 3.} Colour-magnitude diagram for 84 unresolved objects detected
in a background field  1.4 degrees from the center of the
Fornax cluster, observed with HST.

\noindent
{\bf Figure 4.}  $B$-band luminosity functions for the HST sample of NGC 1379
globular cluster candidates (solid histogram), and the unresolved objects
in the background field (dashed histogram).  The vertical dotted line
shows the limit of $\sim$100\% completeness of the NGC 1379 sample.

\noindent
{\bf
Figure 5.}  $I$-band luminosity functions
for the objects in the Fornax background field
(solid histogram), for Galactic stars from the Medium Deep Survey in
17 high latitude fields (dotted histogram), and for galaxies and stars
from the CFRS (dashed histogram).  All luminosity functions are 
normalized to 1.0 mag bins and an area of 4.8 arcmin$^2$. The original 
areas are 4.8 arcmin$^2$ for the Fornax background field, 82 arcmin$^2$ for
the MDS
starcounts, and $\sim 425 $ arcmin$^2$ for the CFRS sample.

\noindent
{\bf Figure 6.}  Colour-magnitude diagram for 365 objects with
$0.3 <$ SHARPness $<0.7$ and $-0.4 <$ ROUNDness $<0.4$ in the CTIO image
of NGC 1379. Representative error bars are shown.

\noindent
{\bf Figure 7.}  Completeness as a function of $B$ magnitude for the CTIO
sample shown in Fig. 6, 
determined by comparison with the HST sample.  The CTIO sample is
$\sim 100$\% complete at $B\sim 21$ and $\sim 50$\% complete at $B\sim 23.5$.

\noindent
{\bf Figure 8.}  $(B-I)$ distributions for objects in the CTIO sample
(solid histogram), and in the CFRS sample (dotted histogram)
for the magnitude range indicated.
The CFRS sample has been normalized to match the CTIO histogram at
$(B-I) > 2.5$.
The dashed histogram is the $(B-I)$ distribution for the CTIO sample of
objects with galactocentric distance $r < 100$ arcsec
($\sim 9$ kpc).

\noindent
{\bf Figure 9.}  $B$ luminosity function from Fig. 4 corrected for
contamination by subtracting the background luminosity function shown as the
dashed histogram in Fig. 4.  The maximum-likelihood best fitting
Gaussian function is shown
as the solid curve, and has $\langle B\rangle =24.95\pm0.30$
and $\sigma_B=1.55\pm0.21$.  

\noindent
{\bf Figure 10.}  Surface density profiles (number per arcmin$^2$)
for objects in the HST sample.
(a) The filled circles are for globular cluster candidates
with $B<25.5$, to which limit the sample is expected to be
both complete and uncontaminated.
The open circles are
for objects with $B>26.5$, which are probably
mostly background galaxies.  The solid horizontal and dahsed lines indicate
the background level derived from the HST background field for $B < 25.5$
and for 
$B > 26.5$ respectively. 
The cluster system can be  traced to a radius
of 80 arcsec ($\sim 7$ kpc).  The average number of clusters observed
in the last 5 bins is $< 4$.
The dip in the number of clusters in the central
$\sim 10$ arcsec appears to be real. (b) Radial profile for the
sample of globular clusters with $B < 25.5$ with a
background of 2.7 objects per arcmin$^2$ subtracted.
The dotted curve is the surface brightness profile of the
underlying galaxy, arbitrarily normalized to match the radial
profile of the cluster system.  Poisson errorbars are shown.

\noindent
{\bf Figure 11.}  Surface density profiles (number per arcmin$^2$) for
the CTIO sample
(filled circles), and the HST sample with $B<25.5$ (solid curve).
Poisson errorbars are shown.  The CTIO surface density profile 
has been increased by 0.74 in $\log N$ 
to match the HST distribution.
Beyond $\sim 100$ arcsec
($\sim 9$ kpc) the
CTIO sample shows no radial gradient.
The solid horizontal line corresponds to a background density of
$\sim 9$ objects per arcmin$^2$.

\noindent
{\bf Figure 12.}  $(B-I)$ distributions for the CTIO sample of NGC 1379
globular cluster candidates with galactocentric radius $<100$ arcsec
(shaded histogram), for the background (CFRS) corrected sample (solid 
histogram),
and for 95 globular clusters in the Milky Way (dashed histogram).

\clearpage
\input{psfig}

\begin{figure}
\psfig{figure=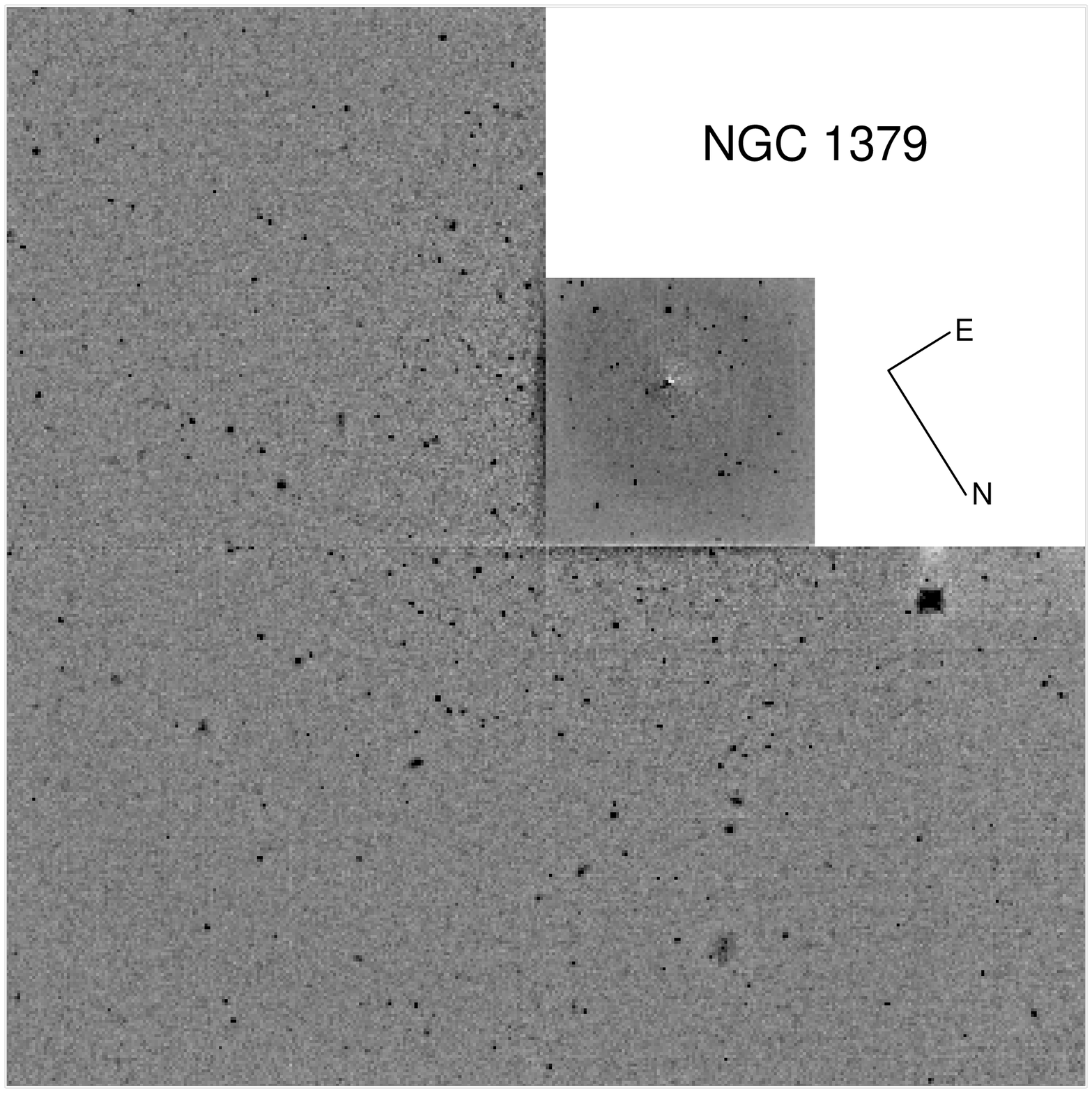,height=7.0in}
\caption{}
\end{figure}

\begin{figure}
\psfig{figure=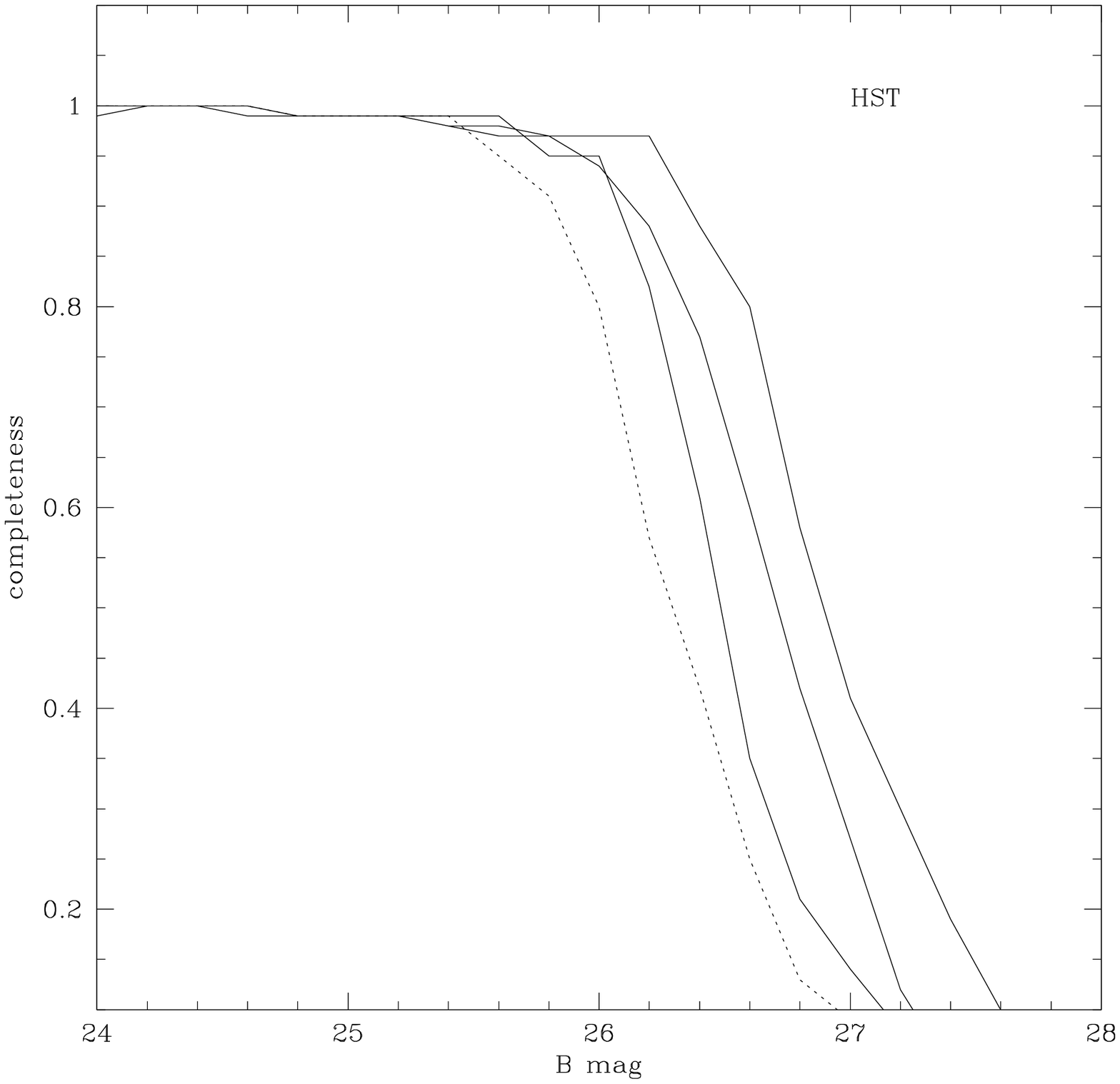,height=7.0in}
\caption{}
\end{figure}

\begin{figure}
\psfig{figure=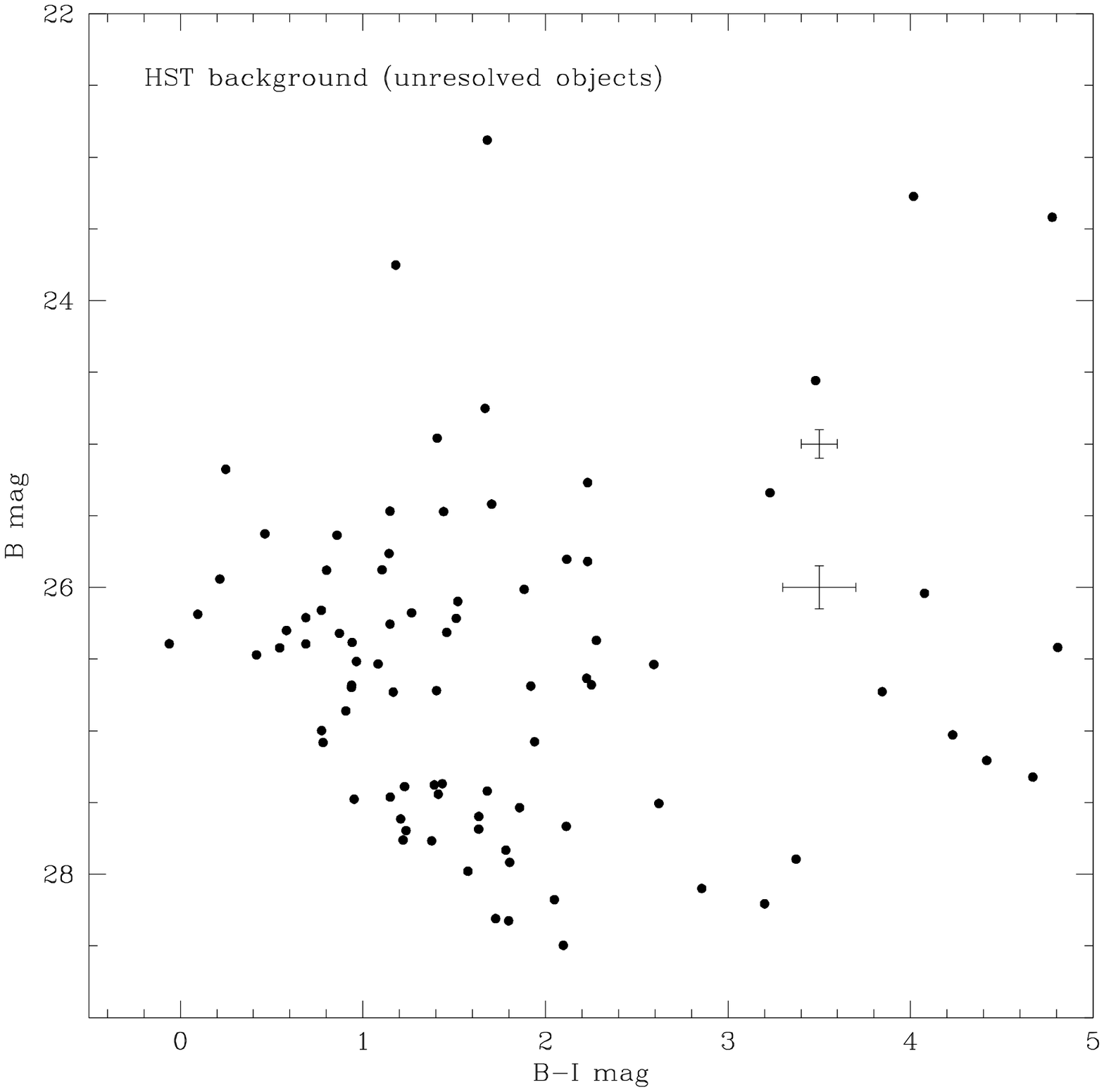,height=7.0in}
\caption{}
\end{figure}

\begin{figure}
\psfig{figure=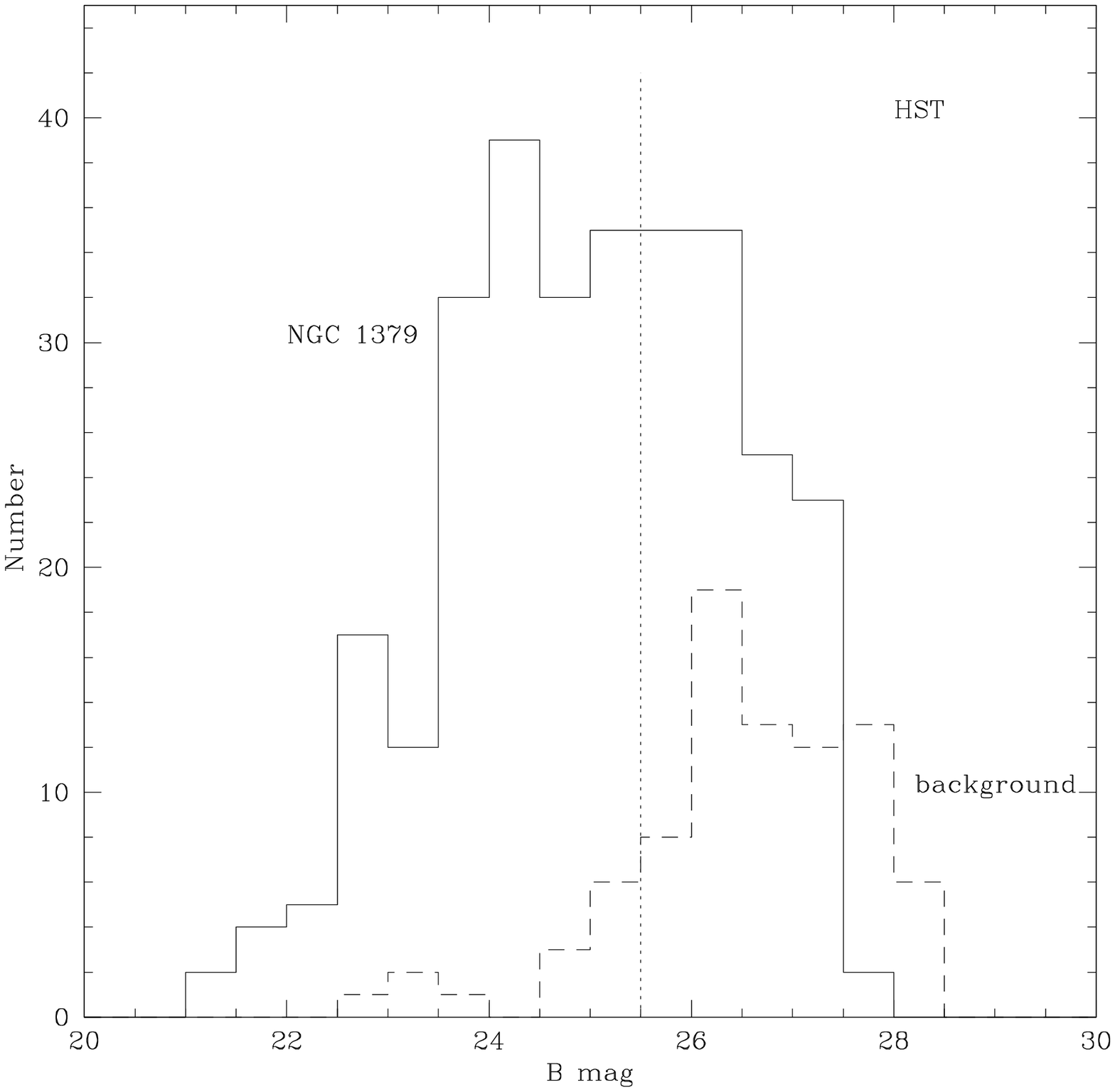,height=7.0in}
\caption{}
\end{figure}

\begin{figure}
\psfig{figure=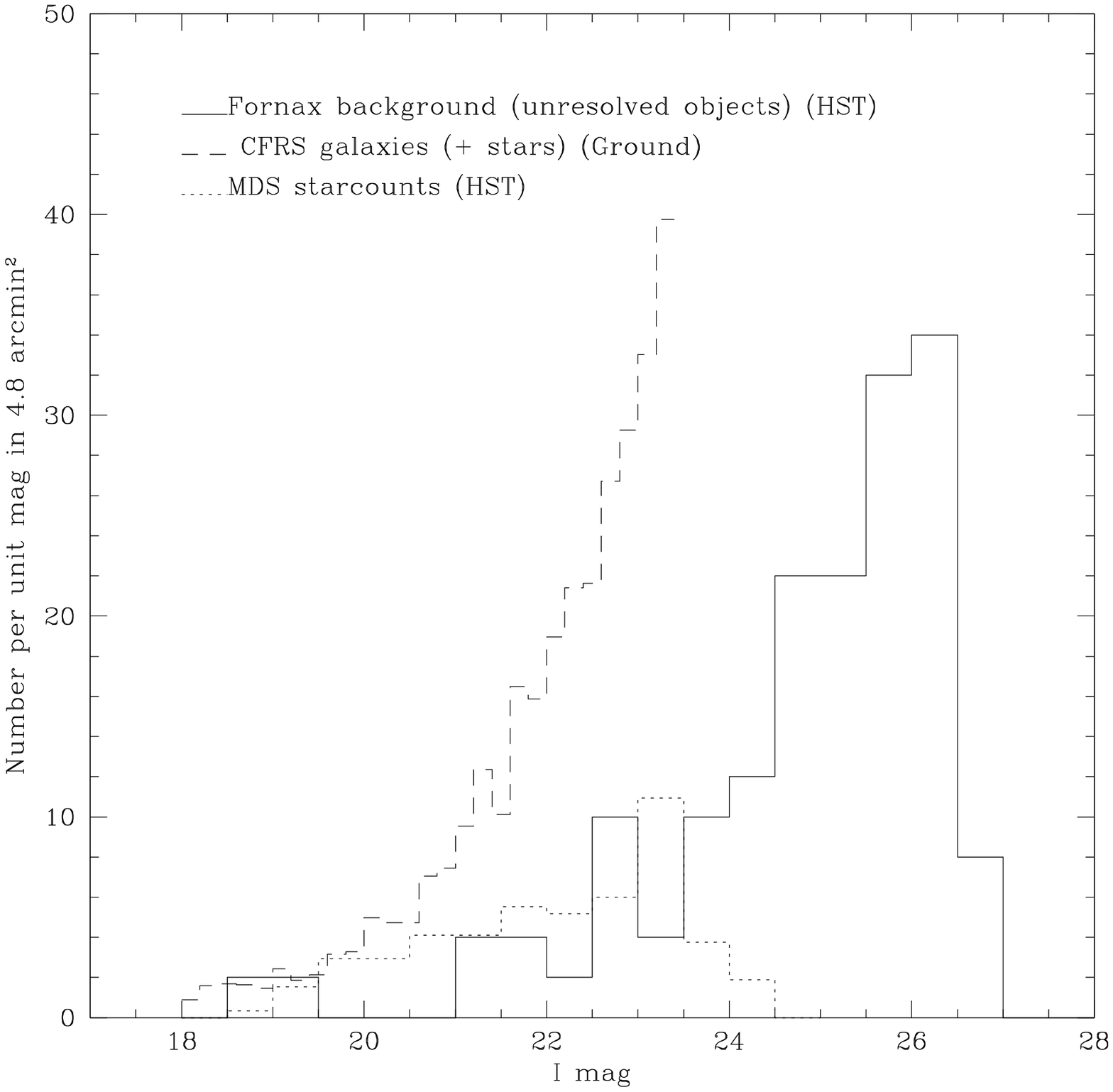,height=7.0in}
\caption{}
\end{figure}

\begin{figure}
\psfig{figure=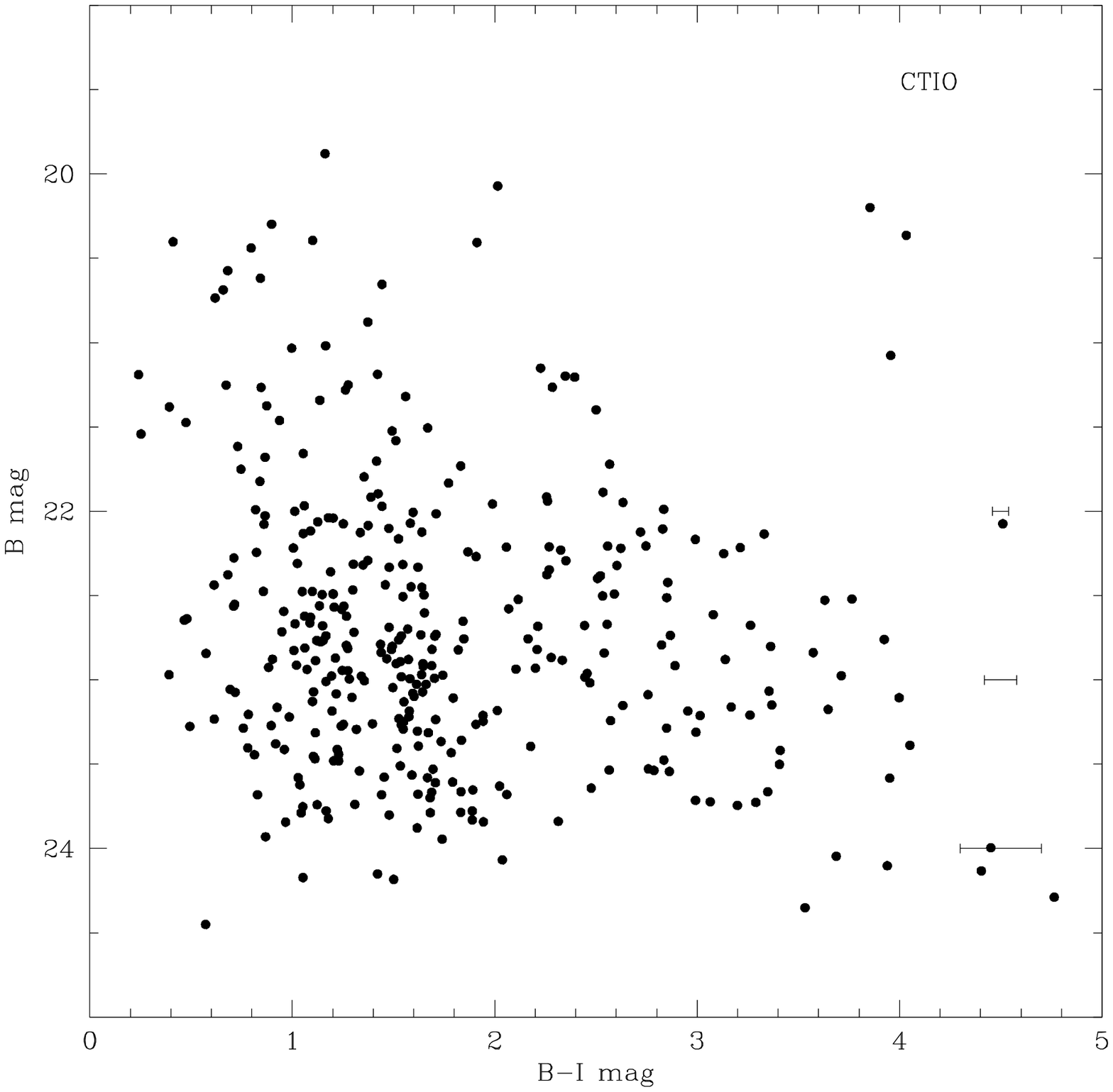,height=7.0in}
\caption{}
\end{figure}

\begin{figure}
\psfig{figure=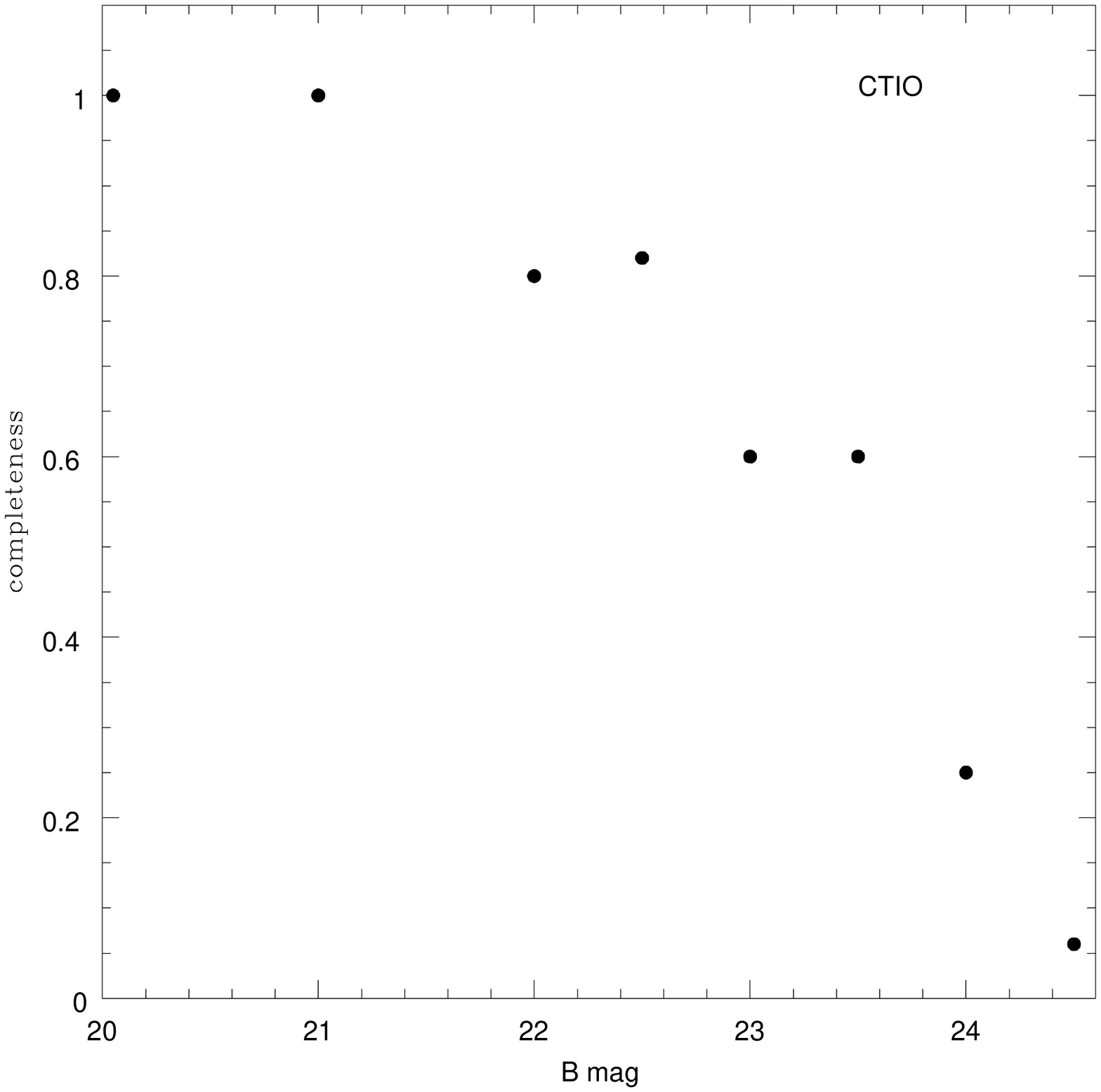,height=7.0in}
\caption{}
\end{figure}

\begin{figure}
\psfig{figure=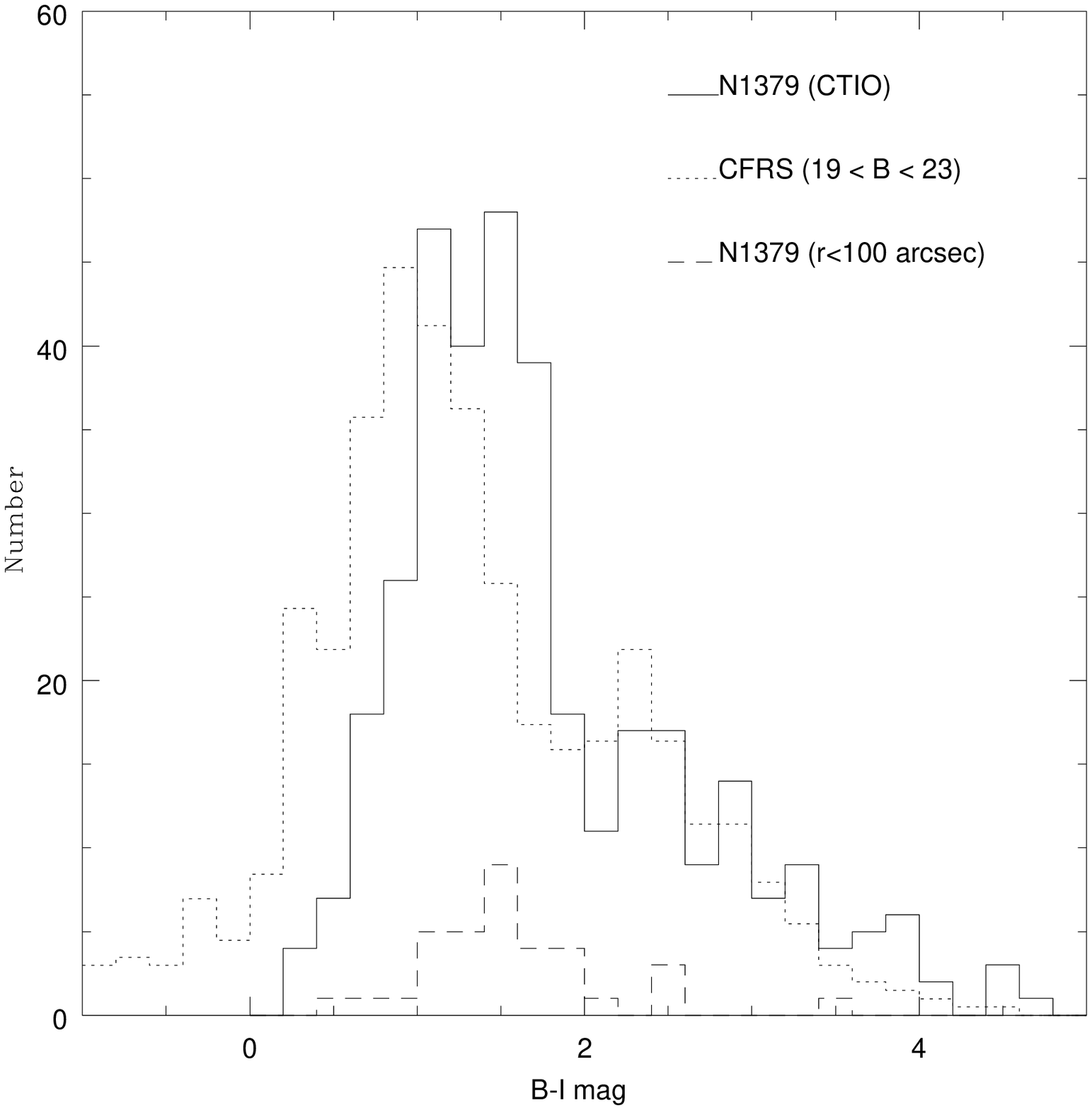,height=7.0in}
\caption{}
\end{figure}

\begin{figure}
\psfig{figure=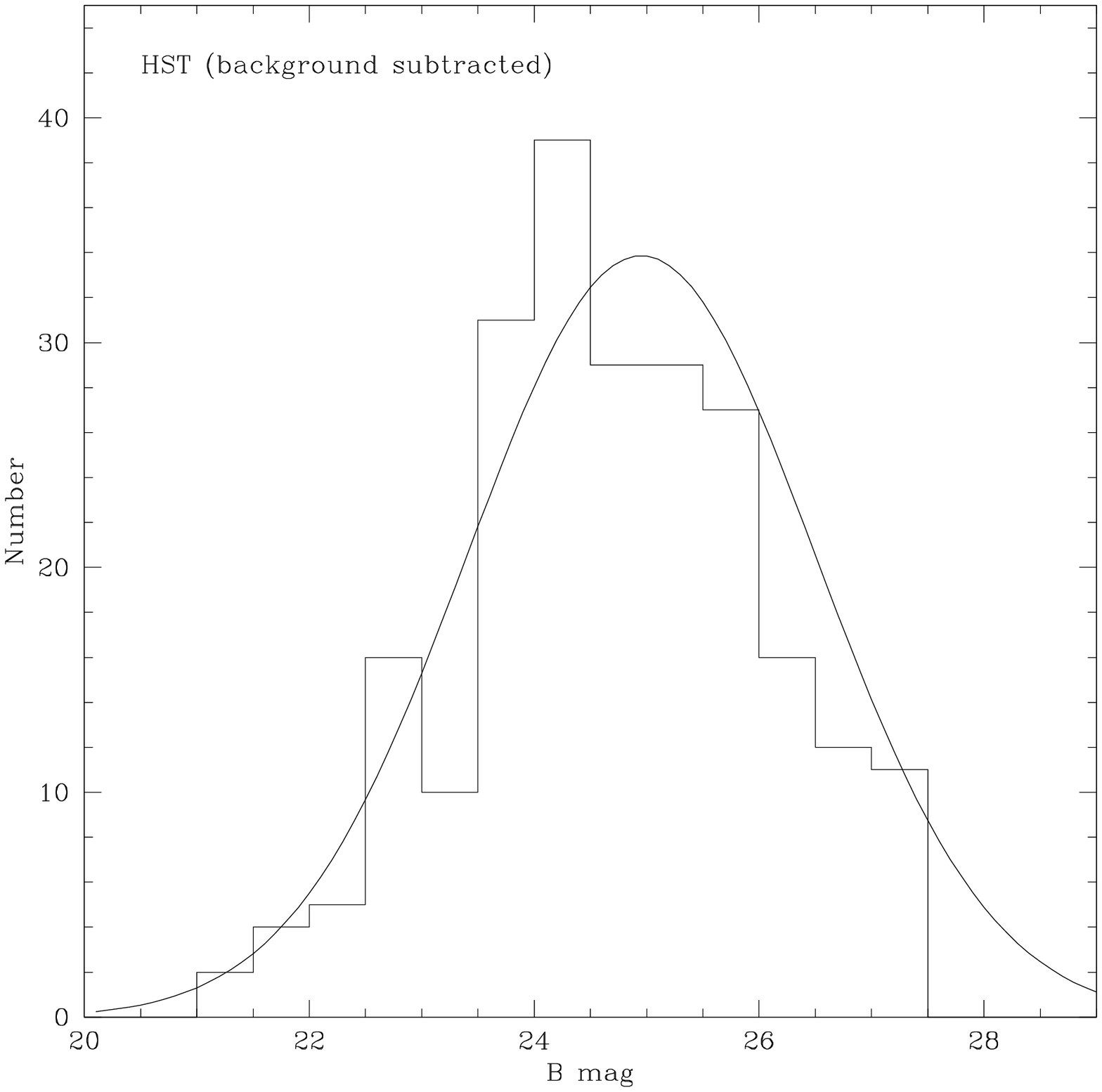,height=7.0in}
\caption{}
\end{figure}

\begin{figure}
\psfig{figure=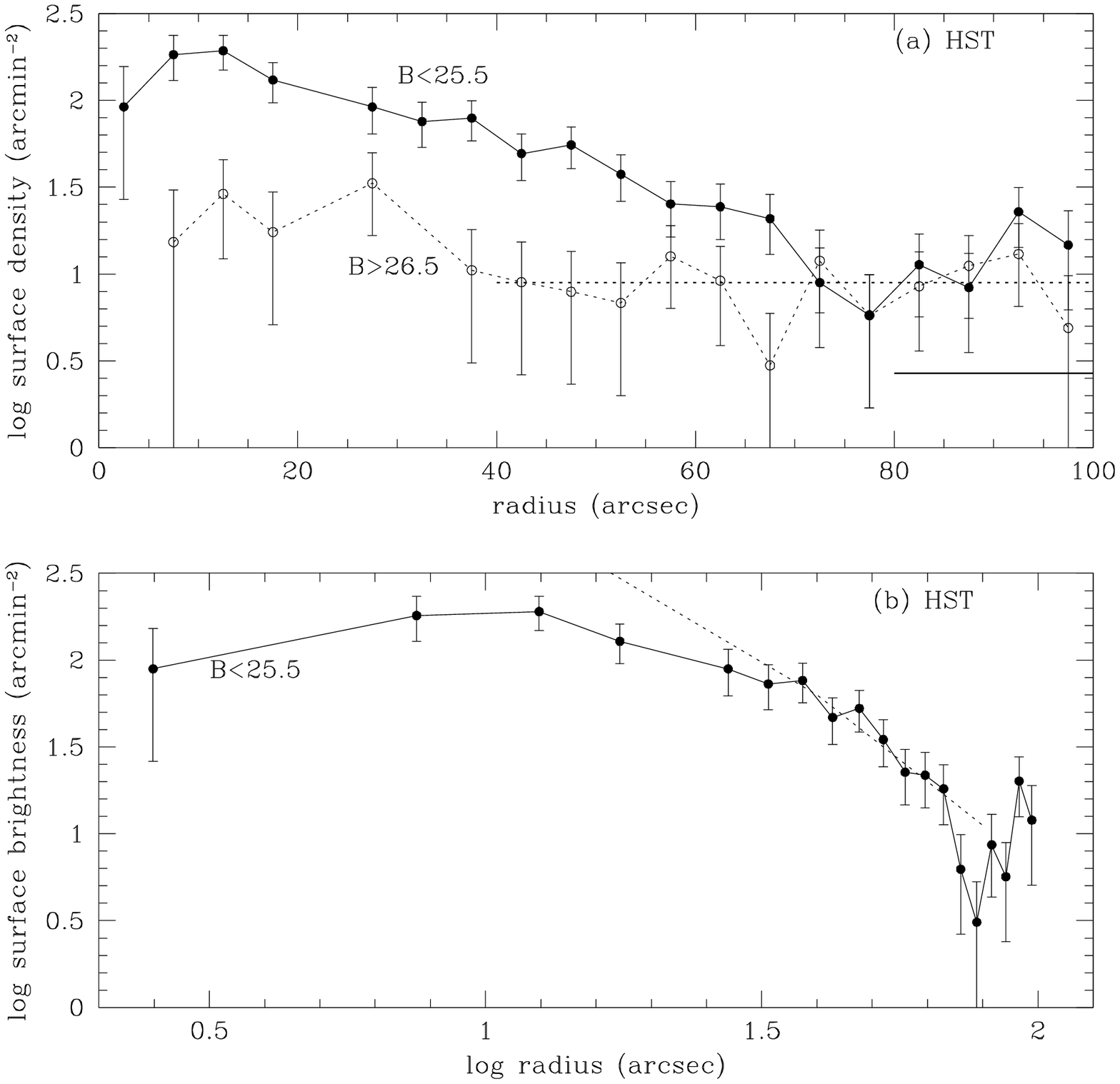,height=7.0in}
\caption{}
\end{figure}

\begin{figure}
\psfig{figure=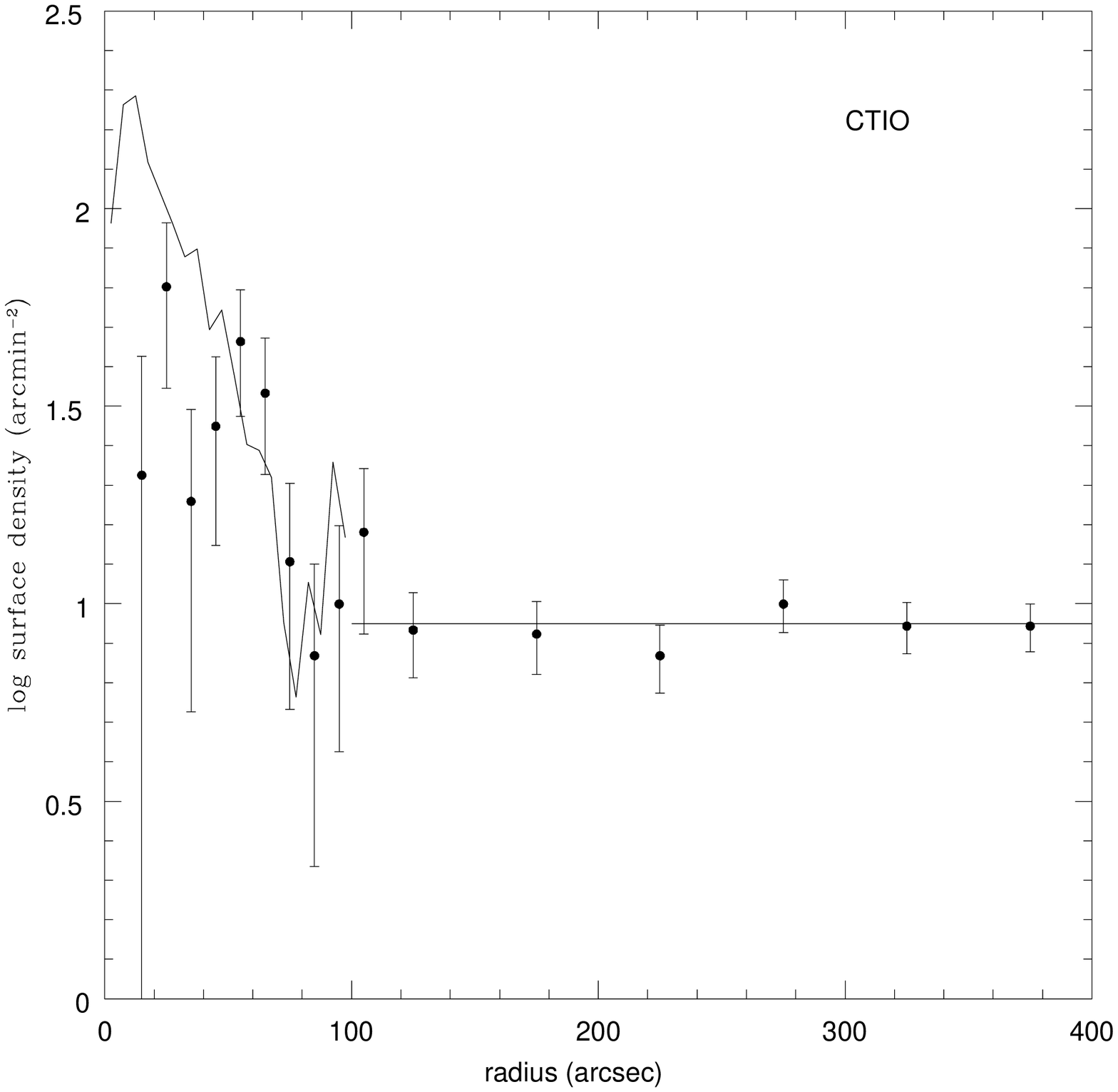,height=7.0in}
\caption{}
\end{figure}

\begin{figure}
\psfig{figure=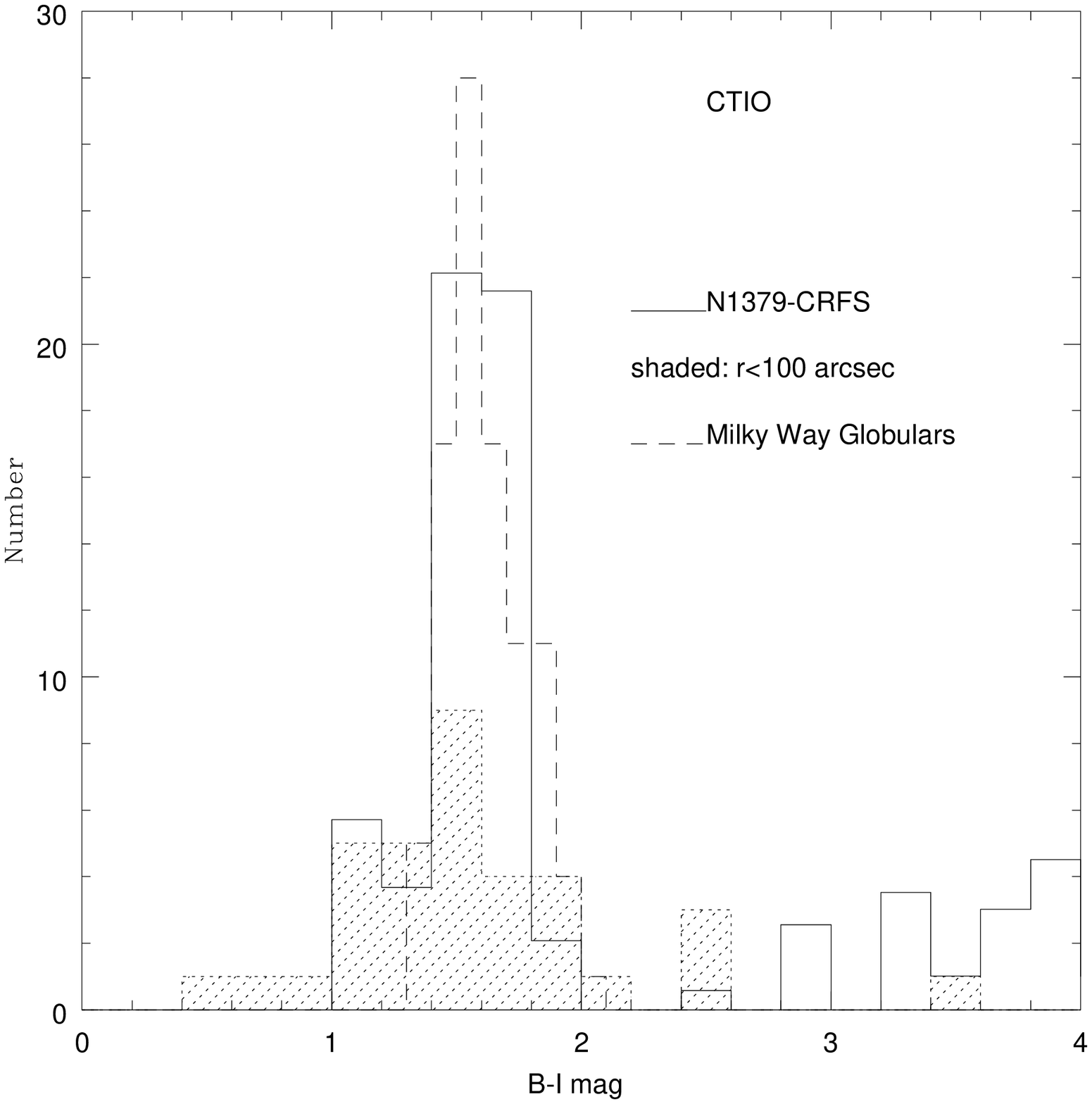,height=7.0in}
\caption{}
\end{figure}

\end{document}